\documentclass[%
 reprint,
 amsmath,amssymb,
 aps,
]{revtex4-2}

\usepackage{graphicx}% Include figure files
\usepackage{dcolumn}% Align table columns on decimal point
\usepackage{bm}% bold math
\usepackage{xcolor}
\usepackage{mathrsfs}
\usepackage[normalem]{ulem}
\usepackage{caption,subcaption}
\def\rg{\textcolor{magenta}}

\def\St{\textrm{St}}

\begin{document}

% \preprint{APS/123-QED}

\title{{Irregular dependence on Stokes number and non-ergodic transport of heavy inertial particles in steady laminar flows}}% Force line breaks with \\

\author{Anu V. S. Nath}
\email{am18d701@smail.iitm.ac.in}
\author{Anubhab Roy}%
 \email{anubhab@iitm.ac.in}
\affiliation{Department of Applied Mechanics,
Indian Institute of Technology Madras, Chennai 600036}

\author{S. Ravichandran}
\email{sravichandran@iitb.ac.in}
\affiliation{ Nordita, KTH Royal Institute of Technology and Stockholm University, Stockholm, Sweden, SE-10691}
\affiliation{Interdisciplinary Programme in Climate Studies, Indian Institute of Technology Bombay, Mumbai 400076}

\author{Rama Govindarajan}
\email{rama@icts.res.in}
\affiliation{International Centre for Theoretical Sciences,
Tata Institute of Fundamental Research, Bengaluru 560089
}

\begin{abstract}
Small heavy particles in a fluid flow respond to the flow on a time-scale proportional to their inertia, or Stokes number $\textrm{St}$. Their behaviour is thought to be gradually modified as $\textrm{St}$ increases. We show, in the steady spatially-periodic laminar Taylor-Green flow, that particle dynamics, and their effective diffusivity, actually change in an irregular, non-monotonic and sometimes discontinuous manner, with increasing $\textrm{St}$. At $\textrm{St} \sim 1$, we show chaotic particle motion, contrasting earlier conclusions for heavy particles in the same flow \cite{wang1992chaotic}. Particles may display trapped orbits, or unbounded diffusive or ballistic dispersion, with the vortices behaving like scatterers in a soft Lorentz gas \cite{klages2019normal}. The dynamics is non-ergodic. We discuss the possible consequences of our findings for particulate turbulent flows.

%The transport of inertial particles in a Taylor Green (TG) vortex flow is shown to have periodic and chaotic dynamics depending on both density ratio, and Stokes number (\textrm{St}) . In the heavy particle limit, only periodic dynamics were observed earlier. However, we show that heavy particles can also have periodic and chaotic dynamics when $\textrm{St}\sim 1$, depending upon their initial conditions.  ($\textrm{St}$).  
\end{abstract}
 
\maketitle

%\tableofcontents

% \section{\label{sec1} Introduction}
Particle-laden fluid flows are common in natural and industrial settings. While sufficiently small particles follow fluid streamlines, finite-sized inertial particles in general do not.
% The suspended tracer particles follow fluid streamlines. However, \rg{inertial particles do not follow fluid elements. Their 
The dynamics of spherical particles of density $\rho_p$ asymptotically greater than the density $\rho_f$ of the ambient fluid, and small enough to be in Stokes flow relative to the fluid, are described by the Maxey-Riley equation \cite{maxey1983equation} in its simplest nondimensional form:
\begin{eqnarray}
%\dot \textbf{v} = \frac{\textbf{u}(\textbf{x})-\textbf{v}}{\textrm{St}} + \textbf{g} \quad \text{with} \quad \dot \textbf{x} = \textbf{v},
% \dot{\textbf{v}} = \frac{\textbf{u}(\textbf{x})-\textbf{v}}{\textrm{St}} + \textbf{g} \quad \text{with} \quad \dot{\textbf{x}}  = \textbf{v},
\dot{\textbf{v}} = \frac{\textbf{u}(\textbf{x})-\textbf{v}}{\textrm{St}} \quad \text{with} \quad \dot{\textbf{x}}  = \textbf{v},
\label{eq1}
\end{eqnarray}
where $\textbf{v}$ is the Lagrangian velocity of the particle, $\textbf{u}$ is the fluid velocity at the instantaneous particle location $\textbf{x}$, and the overdot a derivative in time. The Stokes number $\textrm{St}$ is a measure of particle inertia. 
In this limit, and in dilute suspension, we may neglect the Basset-Boussinesq history, the Saffman lift, effects of added mass and the influence of particles on each other and on the fluid. We also set gravity to zero.

In Taylor-Green (TG) flow, we show a hitherto unsuspected, and often extreme, sensitivity to Stokes number. In some $\textrm{St}$ ranges, `fractions of particles' displaying different behaviour classes are a rough function of $\textrm{St}$. Strangely, in small windows of $\textrm{St}$, all particles can be trapped or moving ballistically. Such behaviour could not have been guessed at earlier.
The model array of TG vortices, providing a time-independent spatially-periodic laminar background flow, has been used previously to study inertial particle dynamics, e.g. by Wang \textit{et al.} \cite{wang1992chaotic} who find particle motion (in the absence of gravity) to be periodic when $\textrm{St} \cdot \textrm{R} \leq 1$ and chaotic when $\textrm{St} \cdot \textrm{R} \geq 1$, where $\textrm{R} \equiv \rho_f/(\rho_p+\rho_f/2)$ quantifies the density ratio between particle and ambient fluid. In particular, they report purely periodic particle dynamics in the heavy particle limit ($\textrm{R}=0$). In contrast, we show that heavy particles of $\textrm{St} \sim 1$ can display chaotic dynamics. We believe the discrepancy is because \cite{wang1992chaotic} investigated only $\textrm{St} \gg 1$. We thus challenge the commonly held view that heavy particle dynamics is often not interesting and easy to predict.

We show that particles display diffusive, periodic or ballistic behaviours depending on $\textrm{St}$ and initial conditions. The resulting dynamics resemble a soft Lorentz gas \citep{klages2019normal} with inertial particles being ``scattered" by the vortices while navigating the stagnation points (SPs). We draw attention to a recent finding which bears some analogy, in a completely different system, of scattering in a soft Lorentz gas \citep{klages2019normal} of irregular variation of diffusion coefficient with system parameters. The behaviour was attributed to the topological instabilities of the periodic orbits. In the large time limit, particles with the same Stokes number but different initial conditions can show different dispersion characteristics, making the transport non-ergodic, in contrast with the usual behaviour of homogeneous and isotropic turbulent flows in which time averages equal ensemble averages. We expect our findings to hold for all flows with periodic coherent structures interspersed with SPs, e.g., Rayleigh-Benard and Taylor-Couette. \\
% Given identical (and zero) initial velocities, we find subsets of initial positions of the particles in two-dimensional space which lead to distinct chaotic or periodic long-time dynamics. These initial positions of identical $\textrm{St}$ particles in the TG flow field form disjoint groups corresponding to periodic and chaotic dynamics (another signature of ergodicity breaking). The fraction of periodic and chaotic particles depends on the Stokes number. \\
We note that our system may be transformed into a form similar to the standard map (see Eq.~(\ref{eq2-whole})), where deterministic diffusion and fractal transport coefficients have been extensively studied \citep{klages2007microscopic}. The discrete ``Chirikov standard map" shows periodic, quasi-periodic, and chaotic dynamics depending on initial condition \cite{chirikov1971research,wood1990arnold}. 
Continuous dynamical systems, such as fluid flow around a periodic array of SPs/obstacles, are also known to induce anomalous (sub-, and super-) diffusion of passive tracers \cite{zaks1996steady, zaks2019subdiffusive, maryshev2020modelling}. Tracers can be chaotic \cite{govindarajan2002universal} in deterministic time-periodic or three-dimensional flows but not in two-dimensional steady deterministic flows like TG flow field. Anomalous diffusion now results instead from an intermediate stage between order and Lagrangian chaos, which arises due to the singularity in the return time of particles near SPs/obstacles. Our system contains a periodic array of SPs in the TG flow field, and our particles are inertial. Thus the emergence of periodic and chaotic dynamics in the present study is due to the interplay of particle inertia and the existence of SPs.

\begin{figure*}
    \includegraphics[width=0.9\textwidth]{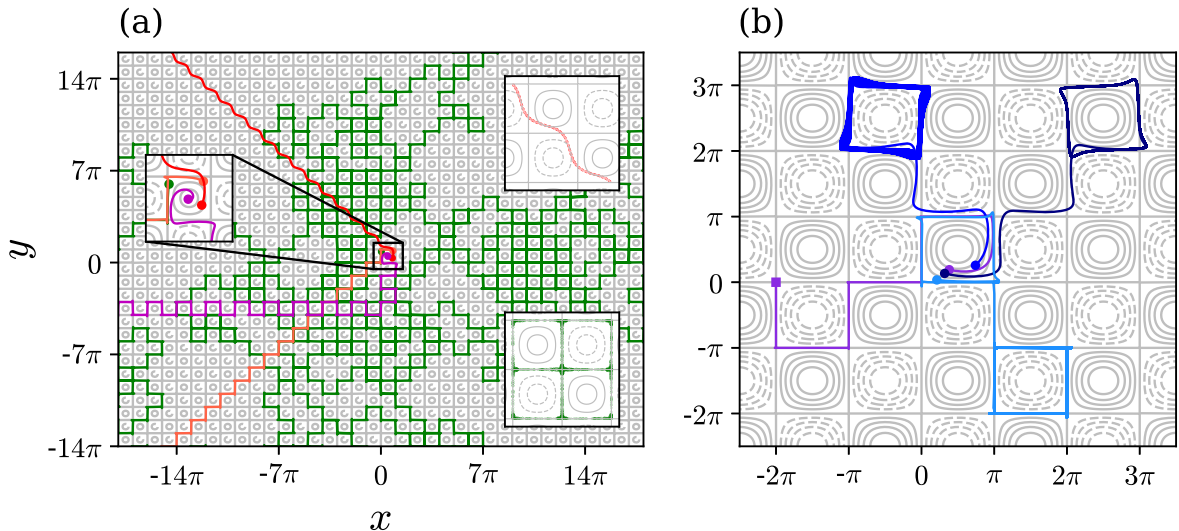}
    \caption{Typical particle trajectories exhibiting ballistic, diffusive and trapped dynamics. The thin grey lines are streamlines of the TG vortex flow. (a) Particles initialised at (i) ($0.11,2.59$) with $\textrm{St} = 1.12$ (green, diffusive), (ii) $(2.6,2.8)$ with $\textrm{St}=0.95$ (orange, diagonal `square ballistic'), (iii) ($2.48,1.07$) with $\textrm{St}=1.83$ (red, diagonal sinusoidal ballistic), and (iv) ($1.5,1.5$) with $\textrm{St} = 1.133$  (magenta, horizontal ballistic). The (left) zoomed inset shows the initial positions by filled circles. The other insets show compactified sinusoidal ballistic and diffusive trajectories 
    % plotted in $x \mod 2\, \pi$ versus $y \mod 2\, \pi$ planes. 
    (b) Trajectories exhibiting trapped
    %, limit cycle and chaotic 
    dynamics at large time. Particles initialised at (i) ($1.0,0.42$) with $\textrm{St} = 1.12$ (navy blue, limit cycle), (ii)  ($2.33,0.81$) with $\textrm{St} = 1.37$ (royal blue, chaotic), (iii)   ($0.664,0.11$) with $\textrm{St} = 0.89$ (azure, trapped in square limit cycle) and (iv) ($1.2,0.6$) with $\textrm{St} = 0.5$ (violet, trapped at the  SP shown by the filled square).}
    \label{fig:trajs}
\end{figure*}

Qualitatively differing trajectories are known in systems unrelated to fluid mechanics \cite[e.g.][]{tanimoto2002phase,geisel1987generic,guantes2001chaos,guantes2003chaotic}. The constant Hamiltonian system in \cite{tanimoto2002phase} displays ballistic, trapped and diffusive trajectories. The effective diffusivity shows fractal dependence on the control parameter (energy). In comparison, our system does not have a constant Hamiltonian (see upcoming paragraph). Nevertheless, we do observe such trajectories (see Fig.~\ref{fig:trajs}) and moreover, a fractal dependence of effective diffusivity with control parameter Stokes number (see the inset in Fig.~\ref{fig:fractions}). 
The equations (\ref{eq1}), for a two-dimensional steady TG flow of stream function $\psi = \sin x \, \sin y$,  written in the rotated coordinates $z = x+y$, $w = x-y$ become
\begin{subequations}
\label{eq2-whole}
\begin{eqnarray}
\textrm{St}\, \ddot{z}+\dot{z} = \sin w~, \\
\textrm{St}\, \ddot{w}+\dot{w} = \sin z~.
\end{eqnarray}
\end{subequations}
% The effect of gravity is neglected here as in \cite{wang1992chaotic}.
This is a dynamical system in four-dimensional phase space with variables $(z, w, v_z, v_w)$, where $\dot{z} = v_z = v_x+v_y$ and $\dot{w} = v_w =  v_x-v_y$. This apparently simple system, equivalent to two coupled damped oscillators with nonlinear, periodic forcing, can nevertheless produce complicated dynamics, as we shall see.
The system is dissipative and equations (\ref{eq2-whole}) may be derived from a time-dependent Hamiltonian $\mathcal{H}(t) = (T + V )\, e^{t/\textrm{St}}$, with
% can generate the governing equations through Euler-Lagrange equations. 
$T = (\dot{x}^2-\dot{y}^2)\, \textrm{St}/2$ and $V = \cos x \, \cos y$, equivalent to the kinetic and potential energy respectively. 
%The unfamiliar form of kinetic energy resembles that of a particle in a \tba{Minkowski plane $\mathbb{R}^{2}_{1}$ \cite{bayrakdar2018time} and may relate to that in a complex plane} (see \cite{bender2011complex}). 
We may relate Eqs.~(\ref{eq2-whole}) to Bateman's time-independent Hamiltonian for dissipative systems \cite{bateman1931dissipative} where the dual system would correspond to dynamics of inertial particles experiencing ``negative" drag (see supplementary material \cite{supplement}, Section~I).
% Since the Hamiltonian mentioned above is explicitly dependent on time, it can not be a conserved quantity. 
Interestingly, as may be verified, the quantity $\mathcal{Q} = T+ V + \frac{2}{\textrm{St}}\, \int_{0}^{t}T\, dt$ is a constant of our dynamics. We use this fact to check the accuracy of our numerical simulations in Fig. ~\ref{fig:trajs} (see supplementary material \cite{supplement}, Section~VII).

% It can be thought of as the dynamics of a single particle starting with different initial conditions since it is assumed that there is no interaction among the particles in the cloud.
A cloud of non-interacting particles is initialized within a ``basic vortex cell" (exploiting the spatial symmetry) of dimensions $\pi \times \pi$, consisting of a central vortex with four SPs at its corners and bounded by separatrices $x=0$, $x=\pi$, $y=0$ and $y=\pi$. For each Stokes number, we initialize $10000$ particles (unless specified otherwise) uniformly in the basic vortex cell and track each particle in time by integrating Eqs.~(\ref{eq2-whole}) using a Runge-Kutta scheme of fourth-order accuracy. The size of the time step was fixed in each case so as to preserve the constant of dynamics $\mathcal{Q}$ 
 upto a maximum percentage error of $0.1\%$, see supplementary material \cite{supplement}, Section~VII.
 %\textcolor{blue}{The results presented here are produced with a time step $dt \leq 10^{-2}$ which is found to be enough to preserve the constant of dynamics $\mathcal{Q}$ as shown in the supplementary material \cite{supplement}, Section~VII. However, much smaller values of $dt$ are used for specific trajectories, especially for chaotic ones, and validated for the repeatability of our results.}. 
% \tba{Where relevant, we average over the dynamics of particles initialised in a uniformly distributed manner}.
%Where relevant, we evaluate the average dynamics of particles, for which the average is taken over particles, which are initialised in the basic vortex cell with uniform distribution. 
The initial velocity of all particles is set to zero (a different choice of initial velocity does not alter the dynamics qualitatively; see \cite{supplement}, Section.~V). We classify trajectories as trapped, diffusive or ballistic according to their dynamics after $10^{4}$ time units, using the squared displacement (SD), mean velocity and other methods (see \cite{supplement}, Section~II). 

At small $\textrm{St}$, after the initial stage in which particles get centrifuged out of the central vortex \cite[e.g.][]{maxey1987gravitational,bec2005multifractal,fiabane2012clustering}, the dynamics is primarily confined along separatrices and SPs. Particles with $\textrm{St}<1/4$ remain confined within the vortex cell where they are initialised, whereas particles with higher Stokes can leak through to neighbouring cells. Linear stability analysis \cite[e.g.][]{taylor1963scientific,levin1961studies,bec2005multifractal}, see also \cite{nath2022transport}, yields the critical value $\textrm{St}=1/4$.
%%%%%%%%%%%%%%%%%%%%%%%%%%%%%%%%%%%%%%%%%%%%%%%%%%%%%%%%%%%%%%%%%%%%%%%%%%
%%%%%%%%%%%%%%%%%%%%%%%%%%%%%%%%%%%%%%%%%
For $1/4 < \textrm{St} \lesssim 0.775$, we find here that particles leave their initial cell but get trapped at SPs in neighbouring cells.
%\rg{\sout{up to any practical simulation time}}.
When $\textrm{St} \gtrsim 0.775$, we find bounded and unbounded particle trajectories that may be periodic or chaotic, depending on St and initial particle position $(x(0),y(0))$. Typical unbounded and bounded trajectories in physical space are plotted in Figs.~\ref{fig:trajs}(a) and \ref{fig:trajs}(b), respectively. 
% Trajectories similar to our diffusive, sinuous-diagonal ballistic have been previously observed (see e.g. \citep[][]{wang1992chaotic}) while horizontal and vertical ballistic trajectories are a novel observation. 
We find unbounded ballistic trajectories of sinusoidal type (for $\textrm{St} \sim 2$) or `square type' (for $\textrm{St} \sim 1$), which follow the separatrices in a zig-zag manner. Similar zig-zag trajectories were also observed in the ballistic settling of inertial particles in the TG vortex \cite{maxey1986gravitational,marchetti2021falling}. The trajectories in compactified physical space ($x \mod 2\, \pi$ versus $y \mod 2\, \pi$), shown in the insets on the right in Fig.~\ref{fig:trajs}(a), highlight the difference between periodic trajectories, which are one-dimensional, and chaotic trajectories, which are space-filling close to the SPs. This indicates a strange attractor. Within the `diffusive set' of initial locations, two particles of identical $\textrm{St}$ will have different chaotic trajectories; however, they asymptotically approach the same attractor in the {\em compactified} phase space.
%\tr{WHAT is a space-filling trapped trajectory?}
%Two particles starting from different initial locations within the diffusive set will have different chaotic trajectories occupying the same region in the compactified space.
%\tr{Similar arguments apply to ballistic trajectories. \rg{????}}
The diffusion of a particle can be normal or anomalous depending on its residence times in the neighbourhood of SPs \cite{zaks2019subdiffusive}, which is, in turn, a sensitive function of the Stokes number.
% \tr{\sout{The trajectory of a diffusive particle in physical space resembles a two-dimensional random walk. However, the important distinction is that.there exists a finite residence time for }}
In Fig. \ref{fig:trajs}(b), we see typical particle trajectories trapped asymptotically in limit cycles, SPs and strange chaotic attractors. The limit cycles and the trapped-chaotic attractors may appear to have similar shapes; however, when observed closely, it can be seen that the periodic limit cycle trajectories draw a closed curve, while the chaotic trajectories fill a certain region around the separatrices without ever repeating the winding. To differentiate between them, evaluating their respective Lyapunov exponents is the most reliable way to do so. Lyapunov exponents measure the rate at which nearby trajectories diverge or converge, and this can be used to identify the presence of chaos. By calculating these exponents, it is possible to determine whether a given trajectory is part of a limit cycle or a chaotic attractor. More instances of trajectories in compactified physical space may be found in the supplementary material (\cite{supplement}, Fig.~5).
% The trajectories trapped in limit cycles can also be smooth or more square type as shown in Fig.~\ref{fig:trajs}(b). As mentioned earlier, in addition to the periodic limit cycles, there exist quasi-periodic trajectories (both open and closed type). For example, the trajectory of a particle trapped in a quasi-periodic limit cycle is shown in Fig.~\ref{fig:trajs}(b). 

% for a given $\textrm{St}$, we find that the particle dynamics depends on the initial position of the particle, and observe a combination of periodic, diffusive and ballistic particle dynamics. 
% Specifically, we find that while particles with $\textrm{St} \lesssim 0.775$ get trapped in fixed points, particles with $\textrm{St} \gtrsim \mathcal{O}(1)$ are trapped in periodic or quasi-periodic limit cycles. Similarly, ballistic trajectories may either be smooth and sinusoidal, or along a series of straight lines running across vortex cells in a zigzag manner. Finally, the trajectory of diffusive particles resembles a random walk.

Our central finding is that bounded and unbounded trajectories may both occur for heavy particles of a given Stokes number and that their initial positions determine their fate. The domain of initial particle locations in the basic vortex cell consists of disjointed regions, each with distinct large-time dynamics. Thus the behaviour of inertial particles in the TG flow is non-ergodic.

\begin{figure*}
    \centering
    \includegraphics[width=\textwidth]{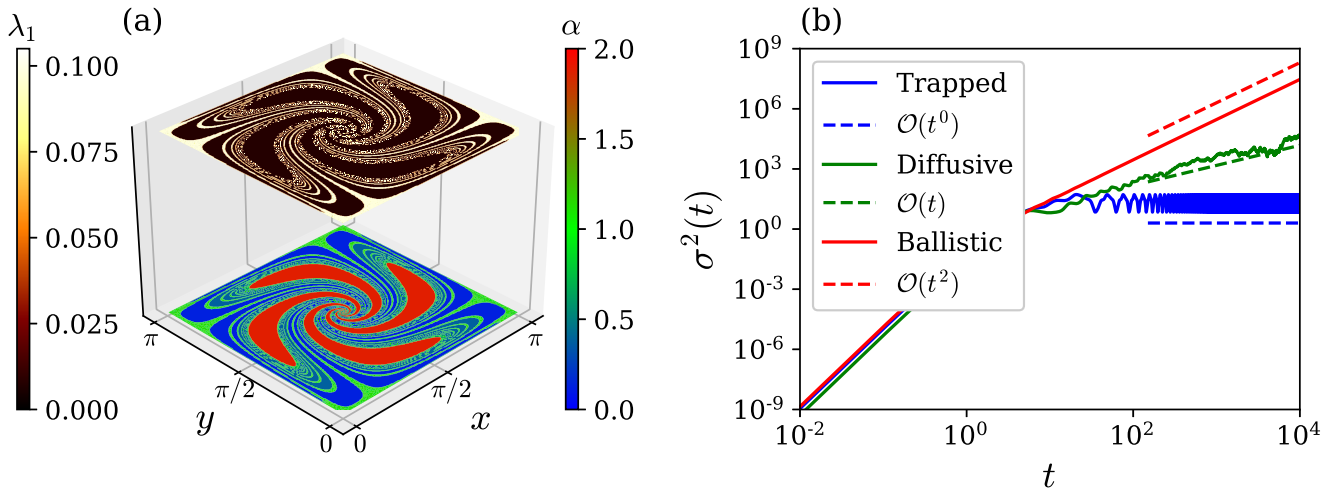}
    \caption{(a) Contour plots, with $316\times316$ particles and $\textrm{St}=1.18$. -bottom: $\alpha$ - such that $\sigma^2(t) \sim t^\alpha$ at large times. -top: $\lambda_1$ - the largest of the `large time Lyapunov exponents'. (b) $\sigma^2(t)$ (solid lines) for particles starting at ($1.5,0.5$), ($0.55,0.15$) and ($0.5,1.0$) and displaying typical ballistic, diffusive and trapped large-time dispersive behaviour respectively; dashed lines show the corresponding asymptotes.}
    \label{fig:alpha_sigma}
\end{figure*}

\begin{figure}
   % \centering
    \includegraphics[width=\linewidth]{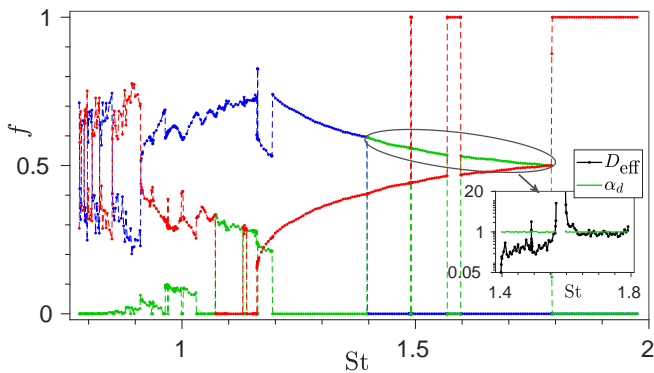}
    \caption{Fraction, $f$, of trapped, diffusive and ballistic particles. The inset shows effective diffusivity $D_{\textrm{eff}}$ and the average $\alpha$ of diffusive particles. %for $1.397 \leq \textrm{St} \leq 1.791$. 
 $\alpha_d \approx 1$ indicates normal diffusion.}
    \label{fig:fractions}
\end{figure}

An example is shown for $\St=1.18$ in Fig. \ref{fig:alpha_sigma}(a) -bottom, showing regions where trapped, diffusive and ballistic trajectories originate in blue, green and red, respectively (we use the same colour code everywhere unless otherwise specified). The trajectories were segregated by evaluating the SD for individual particles. This quantity (SD), denoted by $\sigma^2(t) = \lVert\textbf{x}(t)-\textbf{x}(0)\rVert^2$, where $\lVert \cdot \rVert$ is the Euclidean norm, saturates with time for trapped trajectories, and scales linearly with time for diffusive and quadratically with time for ballistic trajectories. Fig. \ref{fig:alpha_sigma}(a)--bottom plane is thus a contour plot of $\alpha$, 
%\rg{\sout{`the exponent of the squared displacement',}} 
where $\sigma^2(t) \sim t^{\alpha}$ in the limit of $t \rightarrow \infty$.
%%%%%%%%%%%%%%%%%%%%%%%%%%%%%%%%%%%%%%%%%%%%%%%%%%%%%%%%%%%
%Filled contour plot of the exponent $\alpha$ of the squared displacement $\sigma^2(t)$ (such that $\sigma^2(t) \sim t^\alpha$), indicative of the long-time dynamics of $316\times316$ inertial particles with $\textrm{St}=1.18$ initialised uniformly in the vortex cell $(0,\pi)\times (0,\pi)$; trapped (blue), diffusive (green) and ballistic (red) trajectories are possible at this Stokes number.
%%%%%%%%%%%%%%%%%%%%%%%%%%%%%%%%%%%%%%%%%%%%%%%%%%%%%%%%%%%%%%%%%%%
In Fig. \ref{fig:alpha_sigma}(b), we plot the corresponding SD for typical (individual) trapped, diffusive and ballistic trajectories. The SD of trapped particles fluctuates about a mean saturated value, indicating being trapped in a periodic limit cycle, whereas the quadratic increase of SD for the ballistic trajectory is evident. The accurate identification of $\alpha$ is harder for individual diffusive trajectories, e.g., to distinguish between normal ($\alpha = 1$) and anomalous ($\alpha \neq 1$) diffusion. So we segregated ballistic and trapped trajectories (see \cite{supplement}, Section~II) and collectively termed the remaining as `diffusive' (coloured green).
The total fraction of particles that exhibit each kind of behaviour is plotted in Fig. \ref{fig:fractions}. The roughness as a function of Stokes is evident.
%%%%%%%%%%%%%%%%%%%%%%%%%%%%%%%%%%%%%%%%%%%%%%%%%%%%%%
%The fraction ($f$) of trapped, diffusive and ballistic particles plotted for various Stokes number in the range $0.78 \leq \textrm{St} \leq 1.975$. The inset shows the effective diffusivity $D_{\textrm{eff}}$ (defined in the text) of particles exhibiting diffusive trajectories, with $1.397 \leq \textrm{St} \leq 1.791$. Also, the average `exponent of SD' $\alpha$ for all diffusive particles shown in the inset is close to $1$ shows they are showing normal diffusion characteristics.
%%%%%%%%%%%%%%%%%%%%%%%%%%%%%%%%%%%%%%%%%%%%%%%%%%%%%%%%%%%%%%%

We obtain the following regimes for the large-time dynamics, going down in $\textrm{St}$:
(i) For \underline{$\textrm{St} \gtrsim 1.791$}, all particles move ballistically, in a periodic fashion when compactified (see \cite{supplement}, Fig.~8). (ii) For \underline{$1.791 \gtrsim \textrm{St} \gtrsim 1.396$}, depending on the initial location, diffusive or ballistic trajectories are seen. Interestingly, we find two windows, $1.597 \gtrsim \textrm{St} \gtrsim 1.568$ and $1.49116 \gtrsim \textrm{St} \gtrsim 1.4898$, where all particles behave ballistically. 
%Moreover, in the sub-windows $1.59 \gtrsim \textrm{St} \gtrsim 1.569$ and $1.4902 \gtrsim \textrm{St} \gtrsim 1.4898$, all compactified trajectories are periodic. 
%In the remaining \rg{sub-windows}, trajectories show a chaotic Lyapunov spectrum. 
(iii) For \underline{$ \textrm{St} \lesssim 1.396$}, a non-zero fraction of the particles exhibit trapped dynamics. (iv) For \underline{$1.396 \gtrsim \textrm{St} \gtrsim 1.193$}, only trapped and ballistic behaviour are seen, depending on initial conditions. Again, in windows $1.31 \gtrsim \textrm{St} \gtrsim 1.193$ and $1.36 \gtrsim \textrm{St} \gtrsim 1.35$, all  dynamics is periodic. (v) In the interval $1.1602 \gtrsim \textrm{St} \gtrsim 1.072$, ballistic dynamics is not displayed, except by a fraction of particles in the small window $1.139 \gtrsim \textrm{St} \gtrsim 1.13$. (vi) For  $1.072 \gtrsim \textrm{St} \gtrsim 1.03$, unbounded diffusive trajectories are absent, and only trapped and ballistic dynamics are seen. However, all trajectories are periodic for $1.04 \gtrsim \textrm{St} \gtrsim 1.03$. (vii) Finally, for $ \textrm{St} \lesssim 0.775$, particles are attracted to stagnation points and remain trapped there (not shown). For details on dynamics within these windows, see \cite{supplement}, Fig.~8. In Fig.~\ref{fig:fractions}, we used a refinement of $5\times 10^{-3}$ on the Stokes axis and went down to $4 \times 10^{-5}$ where needed. We may have missed smaller windows of Stokes number displaying abrupt changes. Evidence of such smaller windows is visible in the bifurcation diagram shown in \cite{supplement}, Fig.~9.
% From the Lyapunov spectrum we identified that, in this interval, when $1.31 \gtrsim \textrm{St} \gtrsim 1.193$ the dynamics of all particles are periodic (i.e. the trapped and ballistic are periodic); however, when $1.396 \gtrsim \textrm{St} \gtrsim 1.31$, a fraction of particles have chaotic Lyapunov spectrum (i.e. some of ballistic or trapped particles have chaotic Lyapunov spectrum).
All three behaviours (trapped, diffusive and ballistic) co-exist only in some intervals. $\textrm{St} = 1.18$, chosen here for most figures, lies within such an interval.

Incidentally, the disappearance of ballistic dynamics in $\textrm{St} \in (1.072,1.13)\cup (1.139,1.1602)$ may be the reason for the step-like approach of the Mean Square Displacement (MSD)  to ballistic dynamics for condensing particles as their instantaneous Stokes number $\textrm{St}(t) \gtrsim 1$, in \cite{nath2022transport}.
% However, if we closely observe, the shift to ballistic behavior is not a single jump; however, have a small step window (size of the step is inversely related to condensational growth rate) where the dynamics is not ballistic. We conclude that, this happens when the instantaneous Stokes number falls in the above mentioned particular interval. 

The noisiness of the curves in Fig.~\ref{fig:fractions} for $ \textrm{St} \lesssim 1$ is, we believe, intrinsic to the dynamics and not a numerical feature. We claim that this roughness is due to an intrinsic fractal nature and cannot be eliminated by any means. A supporting argument is as follows:
for  $\textrm{St} \in (1.396,1.791)$, we find that the SD averaged over the set of diffusive particle trajectories, $\langle \sigma_D^2(t) \rangle$, scales linearly with time, indicating normal diffusion ($\alpha \approx 1$). Thus, we may evaluate the effective diffusivity in this regime of $\textrm{St}$, as $D_{\textrm{eff}} = \lim_{t \rightarrow \infty} \frac{1}{4}\,\frac{d \langle \sigma_D^2(t) \rangle}{d t}$, plotted as an inset in Fig.~\ref{fig:fractions}, shows roughness. Notice that, in the windows of purely ballistic motion ($1.597 \gtrsim \textrm{St} \gtrsim 1.568$ and $1.49116 \gtrsim \textrm{St} \gtrsim 1.4898$), the effective diffusivity diverges.

The non-ergodicity of our system is reflected in the multimodal distribution of large-time Lyapunov exponents (Fig.~4 in \cite{supplement}), 
which are numerically obtained by applying the algorithm of Wolf \textit{et al.} \cite{wolf1985determining} on Eqs.~(\ref{eq2-whole}). These Lyapunov exponents may also be used to differentiate between periodic and chaotic trajectories.
% The diverse long-time dynamics of particles depending upon their initial conditions motivated us to explore the Lyapunov exponents, \tr{WHAT DOES THIS MEAN:?especially in the long time limit}. 
% \rg{\sout{Lyapunov exponent or Lyapunov characteristic exponent of a dynamical system is a quantity that characterizes the rate of separation of infinitesimally close trajectories in the phase space}}
Eqs.~(\ref{eq2-whole}) have four Lyapunov exponents $\lambda_1 \geq \lambda_2 \geq \lambda_3 \geq \lambda_4$.
%, which together constitute the Lyapunov spectrum as ($\lambda_1,\lambda_2,\lambda_3,\lambda_4$). 
The sign of the exponents indicates the nature of the dynamics (for a detailed discussion, see \cite{supplement}, Section~II C).
%(see \cite{supplement}, Section~II C).
% , e.g., ($0,-,-,-$) indicates a periodic Lyapunov spectrum and ($+,0,-,-$) indicates chaos  \cite{baier1991chaotic, sandri1996numerical,singh2016nature}. 
The dissipative nature of our system of Eqs.~(\ref{eq2-whole}) yields $\lambda_1+\lambda_2+\lambda_3+\lambda_4 = -2/\textrm{St}$. Also, our time-dependent Hamiltonian $\mathcal{H}(t)$ yields a symmetric Lyapunov spectrum as $\lambda_1+\lambda_4 = \lambda_2+\lambda_3 = -1/\textrm{St}$ (see \cite{gupalo1994symmetry,dettmann1996proof,dressler1988symmetry}). Thus, any two Lyapunov exponents are sufficient to estimate the behaviour.
%; for example $\lambda_1$ and $\lambda_2$ alone.
%, as in Fig.~\ref{fig4}. 
The Lyapunov spectrum depends on the initial location of the particles and their Stokes number; e.g. for $\textrm{St} \gtrsim 0.775$, we find nonzero fractions of particles with periodic and chaotic spectra (see Fig.~8 in \cite{supplement}). The dependence of the Lyapunov spectrum on the initial particle locations again indicates non-ergodicity. 

The contour plot of the largest Lyapunov exponent $\lambda_1$,  Fig.~\ref{fig:alpha_sigma}(a)-top, can be seen to mimic
that of $\alpha$ in Fig.~\ref{fig:alpha_sigma}(a)-bottom, showing the correspondence between the dispersion of particles and their dynamical nature at large time. E.g., at $\textrm{St} = 1.18$, ballistic/trapped particles show periodic dynamics and correlate strongly with a periodic Lyapunov spectrum, while diffusive particles show chaotic dynamics and correlate strongly with a chaotic Lyapunov spectrum. There are exceptions in certain windows of Stokes number where ballistic or trapped particles show a chaotic Lyapunov spectrum, and the correspondence breaks down. For example, see the trapped but chaotic trajectory in   Fig.~\ref{fig:trajs}(b)(more in \cite{supplement}, Section~III). Furthermore, all particles with $\textrm{St} \lesssim 0.775$  are attracted to fixed points (SPs) and have a ($+,-,-,-$) Lyapunov spectrum, in the limit $t\rightarrow\infty$. It is well known that for trajectories ending in fixed points, the Lyapunov exponents are the real part of the eigenvalues of the linearized stability matrix about those fixed points \cite{majumdar2001relationships}. Thus, $\lambda_1 > 0$ here indicates just the saddle nature of the fixed point and not any chaotic nature because, otherwise, at least one of the Lyapunov exponents must vanish \cite{haken1983least}. 

Wang \textit{et al.}~\cite{wang1992chaotic} report (i) that the large-time Lyapunov exponents are distributed unimodally and, therefore, (ii) that the large time Lyapunov exponents are independent of initial conditions for any $\textrm{St}$. Our results contradict their findings. %(e.g. the multimodal distributions in Fig.~\ref{fig4})
The main focus of \cite{wang1992chaotic} is on systems with finite density ratios ($\textrm{R}$). In the limit $\textrm{R} \rightarrow 0 $, their limited results only pertain to large Stokes numbers $\textrm{St} \gtrsim 3$, and thus they only find periodic dynamics (open trajectories). We believe their statement is based on their restricted range of study for heavy particles.

In analogy to the conjugacy of the logistic map and the tent map \cite{alligood1997chaos}, we replaced the sinusoids in Eqs. \ref{eq2-whole} with triangular waves (see \cite{supplement}, Section~VI). The latter system
%This is also evidenced by the fact that similar dynamics have been observed in other spatially periodic systems. 
can be solved analytically \cite{supplement} and show similar results.
 
We have shown that heavy inertial particles show a rich tapestry of dynamics even in a very simple two-dimensional cellular flow. The governing dynamical system resembles a billiard system - a viscous soft Lorentz gas, further confirmed by the diverse transport behaviours. The large time dispersion of a particle is shown to be dependent, non-monotonically, and often extremely sensitively, on its inertia (Stokes number) and initial location. The dramatic changes in dispersion with minor changes in $\textrm{St}$ are particularly counterintuitive. When $\textrm{St} \sim 1$, the initial positions in the flow field corresponding to various large time dispersion -- trapped, diffusive and ballistic -- form disjoint groups. The trajectory of a particle starting from one such group can only exhibit one kind of large-time dynamics, indicating ergodicity-breaking. For a range of Stokes numbers, for particles undergoing normal diffusion, the effective diffusivity depends irregularly on $\textrm{St}$, indicating the underlying fractal nature of the dynamics. The `fraction of particles' showing each kind of dispersion shows abrupt transitions as $\textrm{St}$ varies, an underlying feature of fractality. 

Our findings in a TG array are not unique to it. We expect similar behaviour in broad classes of flows containing periodic arrays of vortices and stagnation points. In turbulence, while average dynamics is ergodic, stagnation points and streamlines (separatrices) can last far longer than particle time scales. So individual particles can display non-ergodic behaviour for significant amounts of time. Trapped and chaotic trajectories with positive finite-time Lyapunov exponents might contribute differently to collisions and coalescence. In a cloud, the sampling of moist and dry regimes of the flow may be a sensitive function of Stokes number. We hope the present work will give impetus to test these hypotheses.
% \tr{\sout{Our results may have potential applications in controlling the dispersion of inertial particles in flows via chaotic scattering from ordered layouts of vortices and stagnation points.}}

%%%%%%%%%%%%%%%%%%%%%%%%%%%%%%%%%%%%%%%%
%%%%%%%%%%%%%%%%%%%%%%%%%%%%%%%%%%%%%%%%%%%%%
%%%%%%%%%%%%%%%%%%%%%%%%%%%%%%%%%

\begin{acknowledgments}
SR was supported until June 2022 at Nordita under the Swedish Research Council grant No. 638-2013-9243. Nordita is partially supported by Nordforsk. A.V.S.N. thanks the Prime Minister's Research Fellows (PMRF) scheme, Ministry of Education, Government of India. RG acknowledges support of the Department of
Atomic Energy, Government of India, under project no. RTI4001. A.R. and A.V.S.N. acknowledge
the support of the Complex Systems and Dynamics Group at IIT Madras.
\end{acknowledgments}

% \appendix

% \section{Appendixes}

% \section{A little more on appendixes}

% \subsection{\label{app:subsec}A subsection in an appendix}

%\nocite{*}

\bibliography{ref}% Produces the bibliography via BibTeX.

\end{document}